\documentclass{eas}
\usepackage{graphicx}
\begin{document}
\title{The Wide-field High-resolution Infrared TElescope (WHITE)} 
\author{Denis Burgarella}\address{Observatoire Astronomique Marseille Provence, LAM, denis.burgarella@oamp.fr}
\author{Brice Le Roux}\address{Observatoire Astronomique Marseille Provence, LAM, brice.leroux@oamp.fr}
\author{Maud Langlois}\address{Observatoire Astronomique Marseille Provence, LAM, maud.langlois@oamp.fr}
\author{Gil Moretto}\address{Observatoire Astronomique Marseille Provence, LAM, gil.moretto@oamp.fr}
\author{Thierry Fusco}\address{ONERA, thierry.fusco@onera.fr}
\author{Marc Ferrari}\address{Observatoire Astronomique Marseille Provence, LAM, marc.ferrari@oamp.fr}
\begin{abstract}
The Wide-field High-resolution Infrared TElescope (WHITE) will be dedicated in the first years of its life to carrying out a few (well focused in terms of science objectives and time) legacy surveys.

WHITE would have an angular resolution of $\sim 0.3''$ homogeneous over $\sim 0.7$ sq. deg. in the wavelength range $1 \mu m$ to $5 \mu m$, which means that we will very efficiently use all the available observational time during night time and day time. Moreover, the deepest observations will be performed by summing up shorter individual frames. We will have a temporal information that can be used to study variable objects.

The three key science objectives of WHITE are : 1) A complete survey of the Magellanic Clouds to make a complete census of young stellar objects in the clouds and in the bridge and to study their star formation history and the link with the Milky Way. The interaction of the two clouds with our Galaxy might the closest example of a minor merging event that could be the main driver of galaxy evolution in the last 5 Gyrs. 2) The building of the first sample of dusty supernovae at $z < 1.2$ in the near infrared range ($1.0 - 5.0 \mu m$) to constrain the equation of state from these obscured objects, study the formation of dust in galaxies and build the first high resolution sample of high redshift galaxies observed in their optical frame 3) A very wide weak lensing survey over that would allow to estimate the equation of state in a way that would favourably compete with space projects.
\end{abstract}
\maketitle
\section{Introduction}
From the very first days of astrophysics, astronomers have tried to increase their understanding of the universe by building more and more powerful telescopes. However, there is another way to increase the performances which consists in keeping more or less the same type of instruments but by to install them on an observational site presenting better characteristics in terms of background, transparency, image quality, etc. This quest led astronomers to build telescopes on secluded, remote sites on mountain tops or even to launch them in space. 

However, if we try to summarize these characteristics, keywords are probably cold, dry, stable, high, dark (wavelength dependent) and it seems that there is one site on Earth that qualifies for all of them : Antarctica. To make the best use of the above exceptional characteristics, we propose to build a Wide-field (0.5-degree in diameter) High-resolution ($\sim 0.3$ arcsec using Ground Layer Adaptive Optics) Infrared (from 0.8 to 5 $\mu m$) 2.4-m TElescope that we call WHITE. 

Quite a number of science topics can take advantage of those figures. Burton et al. (2005) reviewed possible science programmes for a 2-m class telescope at Dome C.  However, we will try to show that three well focused key surveys would provide us with the very best and unique science that be carried out within a few years. Of course, beyond the pure science, such surveys would be very helpful to pre-select interesting targets for JWST and ALMA.

\section{Closing the Loop on the Key Science Objectives}
Defining the science objectives, estimating the limits in terms of performance, of time of completion and, in terms of cost is crucial to get the necessary support from the related science community. We will define three key projects that would wonderfully benefit from the observational characteristics available at Dome C.

\subsection{Mapping the Magellanic Clouds in Near- and Mid-InfraRed}

The two Magellanic Clouds are small satellite galaxies of the Milky Way. As so, they provide us with the closest extragalactic objects that we can observe in detail from the South hemisphere. The total LMC has been surveyed with Spitzer (SAGE project : http://sage.stsci.edu/) in the infrared but the angular resolution is only Spitzer's, i.e. 1 - 2 arcsec. We need to go beyond SAGE in the very dense stellar environments that we observe. The two Magellanic clouds are ideally suited to study star formation in a global scale and a survey of the clouds in infrared at a 0.3-arcsec angular resolution will provide a unique database to statistically study star formation outside the Milky Way and resolve stars. The interaction between the Magellanic clouds and the Milky Way is an instance of a minor merger that might have been a common phenomenon in the evolution of large galaxies such as the Milky Way. So, understanding this galaxy-galaxy interaction on a large scale in our neighbourhood is crucial in a cosmological sense.

The infrared wavelength range is particularly interesting to study young stellar formation regions which are embedded into dust clouds. Indeed, since the optical depth of dust decreases with increasing wavelengths (Figure 1), it is possible to detect and study objects that would be otherwise deeply buried into dust and, therefore, undetected at optical wavelengths.

\begin{figure}[htbp]
   \begin{center}
      \includegraphics[width=6cm, height=4cm]{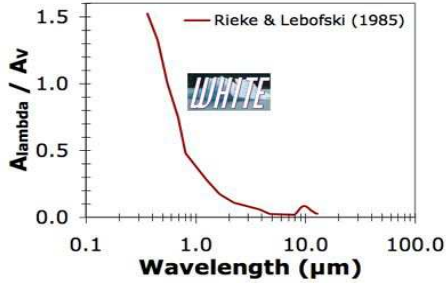}
   \end{center}
   \caption{There is a strong decrease of the dust attenuation with the wavelength down to a few $\mu m$. The WHITE logo indicates the wavelength range that we would like to use.}
\end{figure}

\subsection{Detecting High Redshift Galaxies and Dusty Supernovae at $z < 0.8$}

The high sensitivity in the NIR/MIR opens up the possibility to study very large populations of high redshift galaxies (half a million galaxies per sq. deg. for the shallow night time survey) that will be observed in their optical frame. The observations also offers a very interesting opportunity to detect a new population of dusty supernovae that can hardly be observed in the visible range (e.g. Maiolino et al. 2002). We know that most of the star formation in the universe at $z \sim 1$ is hidden in dust (Figure 2). The infrared is therefore a very promising way to explore the history of the star formation in galaxy and the young phases of the formation of galaxies. This topics undergoes a large interest because of the availability of infrared facilities in the recent years (e.g. Chary et al. 2005, Pozzo et al. 2006, Elias-Rosa et al. 2006).

\begin{figure}[htbp]
   \begin{center}
      \includegraphics[width=6cm, height=4cm]{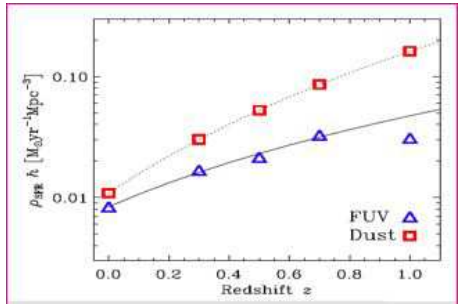}
   \end{center}
   \caption{This figure from Takeuchi, Buat and Burgarella (2005) show that the infrared becomes preponderant at $z \sim 1$ as compared to the local universe.}
\end{figure}

The non detection of dusty supernovae is very likely due to the fact that the dust attenuation is much higher for some supernovae. Maiolino et al. (2002) estimated that the dust attenuation in the V-band ($A_V$) of SN2001db is about 5 magnitudes higher than the average dust attenuation for a more usual sample of supernovae. Astier et al. (2006) stated that "There is no consensus on how to correct for host galaxy extinction affecting high redshift SNeIa ". So, we propose with this survey to decrease its influence by moving to the infrared where the dust attenuation is much lower (by $~ 4$ mag., Figure 1). Going in the infrared will help in the interpretation of SNae light curves by neglecting dust effects .

An additional positive effect would be that light curves in the infrared are flatter for a longer time than in the visible (Krisciunas et al. 2003, Di Carlo 2004).
That means that we have a better opportunity to study supernovae at a higher signal-to-noise ratio in the near infrared than in the visible because the light curve are much more favourable to detect these objets.

Chary et al. (2005) studied supernovae observed with the HST, Spitzer/IRAC and Spitzer/MIPS. They show that an efficient survey of dusty supernovae should be carried out in the NIR/MIR to detect embedded objects but they also stress that high angular resolution is needed to identify supernovae close to the core of galaxies in case of a nuclear starburst.

Figure 3 presents the maximum redshift to detect dusty supernovae estimated in the J, K and L bands. Dashed-dotted horizontal lines show the limiting magnitude for the shallow survey (i.e. using the temporal information) in the three above bands. J and K bands provide about the same maximum redshift of $z \sim 0.4$ in 1-hour exposures within 10 days from the light maximum while the detection of supernovae at maximum would be possible up to z = 0.7 - 0.8, with a slight advantage for the J-band that might be compensented by the fact that the dust attenuation is lower in K than in J.  However, the L-band maximum redshift is $z \sim 0.1 - 0.2$ but with the advantage of observing dustier supernovae.

\begin{figure}[htbp]
   \begin{center}
      \includegraphics[width=10cm, height=7cm]{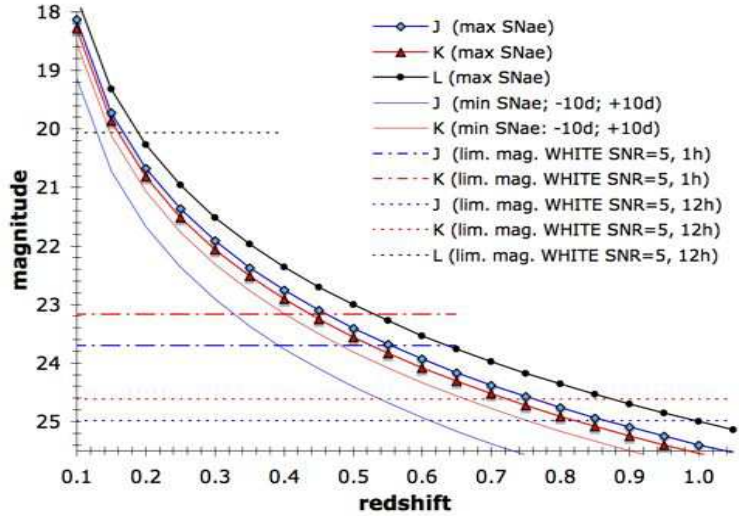}
   \end{center}
   \caption{Redshift range where dusty supernovae (core collapsed) can be detected.}
\end{figure}

Figure 4 shows that at $z = 0.4$, a few dusty SNIa/sq. deg would be observed. However, if we do not restrict ourselves to SNIa but open up to core-collapsed supernovae that are more relevant to study the formation of dust and the star formation of galaxies, a few tens to a few hundreds core collapsed dusty supernovae (within 10 days from the light maximum i.e. 0.2 mag.) can be detected at z = 0.7 - 0.8. This unique sample can only be observed if we can take simultaneously advantage of infrared observations (to look into the dust), of time-dependent observations (supernovae detections), of wide-field observations (for statistics) and for high-resolution (to resolve supernovae from their parent galaxies).

\begin{figure}[htbp]
   \begin{center}
      \includegraphics[width=6cm, height=4cm]{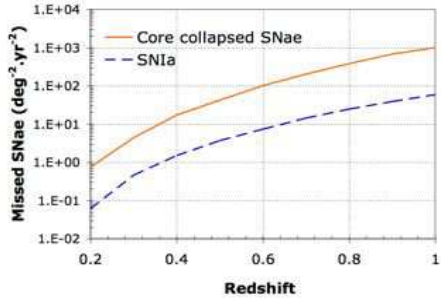}
   \end{center}
   \caption{Quite a large number of dusty supernovae can be detected above $z > 0.5$.}
\end{figure}

Although only a few SNIa supernovae per square degree could be detected at z = 0.4, the number of core collapsed supernovae could reach more respectable numbers up to a few tens. The situation is much more favourable at z = 0.7 - 0.8 where several tens SNIa per square degree and several hundreds core collapsed supernovae  per square degree to build the first large sample of dusty supernovae.

\subsection{Building a Kilo-Degree Survey}
An additional survey would be to use some of the time that might be available if we relax the constraint on the astronomical twilight. A very wide survey over 1000 square degrees in the $K_{dark}$ band would be possible down to $K_{dark} ~ 24$ ABmag. This type of survey would bring a very large sample of galaxies useful, for instance, to constrain the cosmology through the weak lensing. This key project is detailed by Saunders (this volume) and we refer to this paper for a presentation of this very promising topics.

\section{Can we build an instrument able to reach the above specifications ?}

\subsection{Ground Layer Adaptive Optics}
Classic adaptive optics enables large telescopes to provide diffraction limited images, but their corrected field is restrained by the angular decorrelation of the turbulent wave-fronts. However many scientific goals would benefit from a wide and uniformly corrected field, even with a partial correction of about 0.3''.  Ground Layer Adaptive Optics (GLAO) systems are supposed to provide such a correction by compensating the lower part of the atmosphere only and a GLAO correction  appears as a good and adapted system in the specific case of Dome C. A more detailed description of GLAO can be found in Andersen et al. (2006). We performed two independant simulations based on two sets of assumptions. The first set assumed a 4m telescope with a 8x8-actuators deformable mirror from the V-band to the K-band and a turbulence seeing on the ice of 1.1 arcsec. The second set assumed a 2m telescope with a 10x10-actuators deformable mirror from the I-band to the K-band and a turbulence seeing of 1.9 arcsec that are carried out using PAOLA (Jolissaint, Veran, Conan 2006). The first simulations are relatively optimistic while the second one are more pessimistic. The number of actuators does not pose any strong technological constraints. These simulations are encouraging : the gain in encircled energy is still larger than 6 at a radius of 30 arcmin in the red and infrared wavelength ranges (Burgarella et al. 2007). 

The study and development plan of the Ground Layer Adaptive Optics is therefore a priority. After the initial simulations carried out in 2007, we need to test the concept on an antarctized optical bench during 2008 and part of 2009. The phase is necessary to check that the GLAO concept allows to reach the requirements in the lab. Then, it will be necessary to develop a prototype that will be tested on the sky before shipping the prototype to Dome C where it could be tested on a small telescope (e.g. IRAIT). Finally, if the tests are positive the telescope + GLAO + instrumentation would be on the sky in the window 2012 - 2014 to start building the key-project surveys. The completion of this phase should occur in the 2015 timeframe.

\subsection{Optical Designs of a 2.5-m Telescope}

Two types of telescopes are under study at the Observatoire Astronomique Marseille Provence that would allow observing in the NIR+MIR over a wide field of view while not degrading the image quality above 0.3 arcsec. 

\begin{itemize}
\item A first one is the Three-Reflection Telescope (TRT) designed by Lemaitre et al. (2004). The size of the spot is still 0.25 arcsec RMS at a radius of 1 degree, which is well within our specifications. Two down-scaled models of this telescope already exist at IAS, Frascati and at OAMP/LAM, Marseille.
\item Another option would be an off-axis design adapted from space telescope designs (e.g. Moretto et al. 2004) which bears several advantages: no obstruction of the beam by the secondary mirror, no diffraction / emission from the secondary support structure to degrade images / sensitivities. This type of design is studied by G. Moretto. 
\end{itemize}

The feasibility of manufacturing such mirrors for both designs are well understood by several optical manufacturers such as REOSC (Paris-France) and Mirror Lab (Tucson, Arizona, USA) in particular for the 2m M1. In the case of the smaller aspherical mirrors - from 0.43 to 1.1m - we are confident that such mirrors can be made with active vase mirror technology pioneered by Lemaitre.

A concept for an instrument is presented by Mora et al. in this volume.

\subsection{Estimating the performances}

Estimating the performances of WHITE is crucial to evaluate the scientific interest of this project. To compute the performances presented in Figure 4, we assume a 2.4m primary mirror, a night time IR background from Walden et al. (2005), a day time IR background (beyond the K-band) has been estimated by interpolating between the K-band background quoted by Gillett (August 1995) : 9.9 mag/arcsec2 and ~ 35 Jy/arcsec2 at 11.5m (Storey et al. 1998). The temperature of the telescope at 210K (emissivity set to 3\%), a pixel scale of 0.15'', JWST-type detectors (1 - 5 $\mu m$), a Ground Layer AO, integration over 2 * FWHM (EE = 80\%) to optimize the SNR. 

\begin{figure}[htbp]
   \begin{center}
      \includegraphics[width=12cm,height=5cm]{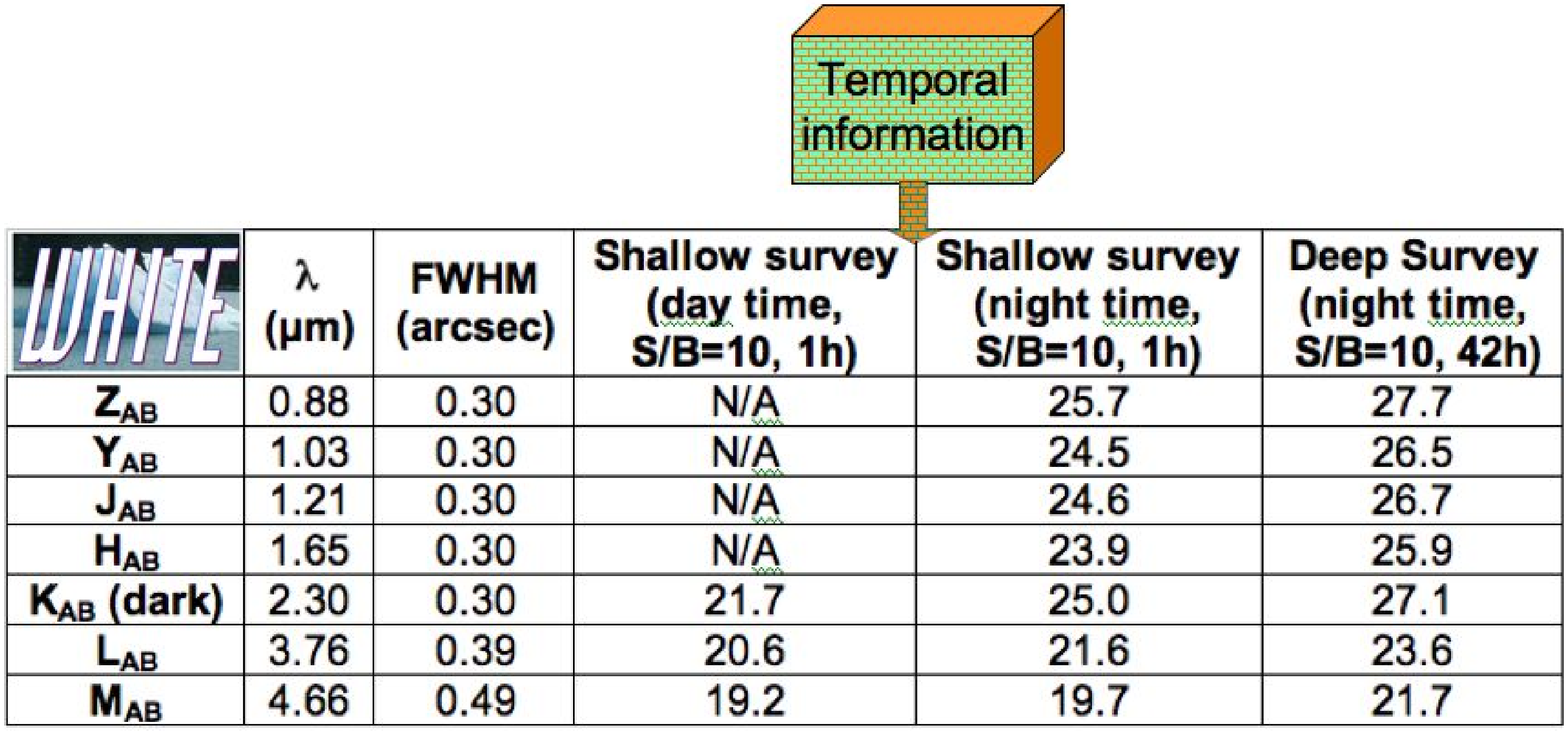}
   \end{center}
   \caption{Limiting magnitudes of WHITE for day time and night time shallow observations (with some temporal information) and deep night time observations.}
\end{figure}

The angular resolution is set by the GLAO device up to the K-band and by the limit of diffraction at longer wavelength. Each individual image of the shallow survey will permit to have a very important temporal information that will be used to track time-dependent phenomenons such as variable stars or supernovae. 

On the other hand, by summing up all the individual exposures, we will be able to reach much deeper sensitivies. Here, we assume a total exposure time of 42h in each band but the total exposure time varies from 42 hours at the shorter wavelength to more than 160 hours in J.

\section{Conclusion}
We present in this paper a concept and the science objectives for the project WHITE that would observe in the approximate range $1 - 5 \mu m$ using a GLAO device to reach an angular resolution of $\sim 0.3$ arcsec over a field of fiew of $\sim 0.75$ sq. degree. It is important to notice that WHITE will efficiently use all the observing time by carrying out observations beyond $\sim 2 \mu m$ even during daytime. To optimize WHITE, we define a few scientific key projects that will be carried out in a limited (2 - 3 years) time.

We must note, however, that other similar projects have been are are proposed like AMIDST (Epchtein 2004 ; Le Bertre, Vauglin \& Epchtein 2006), PILOT (Storey et al. 2007) and a smaller telescope (IRAIT, Tosti 2007) is now at Dome C that will permit to characterize the site from astronomical observations in the wavelength range ($2 - 28 \mu m$). WHITE moves along the same way as the previous ones. Because of environmental and funding issues, a cooperation between several institutes around the World is mandatory.


\begin{thebibliography}{99}
\bibitem[2006]{} Andersen D.R., Stoesz J., Morris S., et al. 2006, astro-ph/061009
\bibitem[2006]{} Astier P., Guy J., Regnault N. et al. 2006, A \& A 447, 31
\bibitem[2006]{} Burgarella D., Ferrari M.,  Fusco T., et al. 2007, EAS Publications Series, Vol. 25
\bibitem[2005]{} Burton M.G., Lawrence J.S., Ashley M.C.B., et al. 2005, PASA 22, 199
\bibitem[2005]{} Chary R., Dickinson M.E., Teplitz H.I., et al. 2005, ApJS 635, 1022
\bibitem[2004]{} Di Carlo E., 2004, MmSAI 75, 150 
\bibitem[2006]{} Elias-Rosa N., Benetti S., Cappelaro E., et al. 2006, MNRAS 369, 1880
\bibitem[2004]{} Epchtein N. 2004,  EAS Publicaions Series, Vol. 14, 193
\bibitem[2006]{} Jolissaint L., V\'eran J.-P., Conan R. 2006, JOSA A 23, 382
\bibitem[2003]{} Krisciunas K., Suntzeff N.B., Candia, P. et al. 2003, AJ 125, 166
\bibitem[2006]{} Le Bertre T., Vauglin I ., Epchtein N. 2006, EAS. Publications Series, Vol. 25, 179
\bibitem[2004]{} Lemaitre G.R., Montiel, P., Joulie, P., et al. 2004, SPIE 5494, 426 
\bibitem[2002]{} Maiolino R., Vanzi L., Mannucci F., et al. 2002, A\& A 389, 84
\bibitem[2004]{} Moretto G., Langlois, M., Ferrari, M. 2004, SPIE 5487, 1111
\bibitem[1998]{} Storey J.V.W., Ashley M.C.B., Burton M.G., Pet al. 1998,  SPIE 3354, 1158
\bibitem[2007]{} Storey J.W.V., Ashley M.C.B., Burton M.G., et al. 2007,  EAS Publications Series, Vol. 25, 255
\bibitem[2006]{} Pozzo M., Meikle W.P.S., Rayner J.T., et al. 2006, MNRAS 368, 1169
\bibitem[2004]{} Takeuchi T.,T., Buat V., Burgarella D. 2005, A \& A 440, L17, 147
\bibitem[2007]{} Tosti G. 2007,  EAS Publications Series, Vol. 25, 209
\bibitem[2005]{} Walden V.P., Town M.S., Halter B., et al. 2005, PASP 117, 300
\end{thebibliography}
\end{document}